\documentclass[fleqn,twoside]{article} 
\usepackage{epsf,multicol,ifthen}
\usepackage{ujp}
\usepackage[cp1251]{inputenc}
\usepackage[english,russian]{babel}
\usepackage{amstext}
\usepackage{amssymb}
\usepackage{cite}
\usepackage{graphicx}
\nazva{PECULIARITIES OF THE LIGHT\\ ABSORPTION AND EMISSION BY
FREE\\
ELECTRONS IN MULTIVALLEY SEMICONDUCTORS}%

\udk{535.341}

\nazvacol{PECULIARITIES OF THE LIGHT ABSORPTION}%

\avtor{P.M. TOMCHUK}%
\avtorcol{P.M. TOMCHUK}%

\inst{Institute of Physics, Nat. Acad. Sci. of Ukraine}%
\adr{(46, Nauky Prosp., Kyiv 03028, Ukraine)}

\begin{document}           
\setcounter{page}{681}%
\maketitl                 
\begin{multicols}{2}
\anot{%
General expressions are obtained for the coefficient of light
absorption by free carriers as well as the intensity of the
spontaneous light emission by hot electrons in multivalley
semiconductors. These expressions depend on the electron
concentration and electron temperature in the individual valleys.
An anisotropy of the dispersion law and electron scattering
mechanisms is taken into account. Impurity-related and acoustic
scattering mechanisms are analyzed. Polarization dependence of the
spontaneous emission by hot electrons is found out. At
unidirectional pressure applied or high irradiation intensities,
the polarization dependence also appears in the coefficient of
light absorption by free electrons.}

\section*{Introduction}

The phenomena of the light absorption and emission by free carriers
have been studied for years, and it seems likely that a discovery of
new effects can hardly be expected here. However, this is not true
in the case of multivalley semiconductors. The peculiarities of the
mentioned phenomena in such semiconductors are related to (i) a
sharp anisotropy of the dispersion law for electrons, (ii) the fact
of the electron filling of several valleys and, finally, (iii) an
anisotropy of scattering mechanisms. It is known that a ``third
body'' is required in the act of the photon emission or absorption
by a free electron. This ``third body'' provides the energy and
momentum conservation during collisions. Impurity atoms, lattice
oscillations (phonons), or boundaries can serve for it. By this, the
influence of scattering mechanisms (including their anisotropy) on
the absorption and emission processes can be explained.

In the thermodynamic equilibrium state, the quantity of photons absorbed by
free electrons is equal to that of emitted photons. The detailed balancing
principle is therefore valid. If the photon quantity exceeds that in
equilibrium state, i. e., a semiconductor is irradiated with an external
electromagnetic field, the photons are absorbed by free carriers. If the
external electromagnetic field is absent, and the electron gas is heated
(e.g., by a constant electric field), the process of light emission by free
electrons occurs.

A new effect related to the light emission by free electrons in multivalley
semiconductors, comparing to single-valley semiconductors, is the appearance
of the polarization dependence of the emitted light intensity. The same
dependence can also appear in absorption if the irradiation intensity of a
multivalley semiconductor is high enough.

There are various methods that allow finding the absorption and
emission by free carriers. In our opinion, the most convenient
method is to use the kinetic equation, in which the influence of the
electromagnetic field on the free carriers scattering mechanism is
taken into account. The convenience of this method lies in that one
may derive the expression for the absorption by free carriers both
in classical and quantum cases in a single approach. Besides
absorption, this method allows also finding the wave-field-induced
emission by free carriers. From here, one may obtain the spontaneous
emission by free carriers through definite formal substitutions. We
have used such an approach in [1]. In that paper, we have
investigated the mechanism of acoustic scattering in detail, and
outlined a model for the impurity scattering. However, no analytic
expressions for both classical and quantum absorption under the
impurity scattering in multivalley semiconductors have been derived.
The same is true also for the emission regularities. That is why we
pay the principal attention in this paper to the study of the
situations where the impurity scattering dominates.

\section{Collision Integral of Electrons with Ions in the Presence of an
Electromagnetic Wave}

We consider the multivalley semiconductors like $n$-Ge and $n$-Si. The Hamiltonian
of electrons, which populate one of the conduction band valleys, can be
written in the principal axes of the mass ellipsoid in the presence of an
electromagnetic field as follows:
\begin{equation}
\label{eq1}
\hat {H} = {\sum\limits_{\alpha = 1}^{3} {{\frac{{1}}{{2m_{\alpha}
}}}\left( {\hat {p}_{\alpha}  - {\frac{{e_{0}}} {{c}}}A_{\alpha}}
\right)^{2} + {\sum\limits_{i = 1}^{N} {U\left( {\vec {r} - \vec {R}_{i}}
\right).}}} }
\end{equation}
In Eq. (\ref{eq1}), $m_{\alpha} $ are the principal components of
the mass tensor, ($m_{x} = m_{y} \equiv m_{ \bot},\,m_{z} =
m_{\|})$, $\hat {p}_{\alpha} $ is the $\alpha {\rm th}$ component
of the momentum operator, $c_{0}$ is the electron charge, $c$ is
the light velocity, $A_{\alpha} $ is the $\alpha {\rm th}$
component of the vector potential of electromagnetic field, $N$ is
the number of ions in volume $V$, $U\left( {\vec {r} - \vec
{R}_{i}} \right)$ is the interaction potential of an electron with
an ion ($\vec {r}$ is the electron coordinate, $\vec {R}_{i} $ is
the coordinate of the $i$th ion),
\begin{equation}
\label{eq2} U\left( {\vec {r}} \right) = {\frac{{e_{0}^{2}}}
{{\varepsilon_{0} r}}}e^{{ -}{r / r_{\rm D}}} ,
\end{equation}
$\varepsilon _{0} $ is the static dielectric constant, and $
r_{\rm D}$ is the Debye radius.

We set the vector potential $\vec {A}$ in a form
\begin{equation}
\label{eq3} \vec {A} = \vec {A}^{\left( {0} \right)}\cos \omega t.
\end{equation}
In formula (\ref{eq3}), $\vec {A}^{(0)}$ is a constant vector,
$\omega $ is the wave frequency.

The electron wave function in the field of an electromagnetic wave but
without scattering centers, is determined from the Schr\"{o}dinger equation
\begin{equation}
\label{eq4} i\hbar {\frac{{\partial}} {{\partial t}}}\psi _{\vec
{p}}^{\left( {0} \right)} = \hat {H}^{\left( {0} \right)}\psi
_{\vec {p}}^{\left( {0} \right)} \equiv {\sum\limits_{\alpha =
1}^{3} {{\frac{{1}}{{2m_{\alpha} }}}\left( {\hat {p}_{\alpha}  -
{\frac{{e_{0}}} {{c}}}\vec {A}_{\alpha}} \right)^{2}\psi _{\vec
{p}}^{\left( {0} \right)}}}
\end{equation}
\noindent and equals:
\[
 \psi _{\vec {p}}^{\left( {0} \right)} = {\frac{{1}}{{\sqrt {V}}} }\exp
\left( {{\frac{{i}}{{\hbar}} }\vec {p}\;\vec {r}} \right)\times
\]
\[
\times\exp {\left\{ { - {\frac{{i}}{{\hbar}} }{\int\limits_{0}^{t}
{dt^{ \prime} {\sum\limits_{\alpha = 1}^{3}
{{\frac{{1}}{{2m_{\alpha}} } }\left( {p_{\alpha}  - {\frac{{e_{0}
}}{{c}}}} \right)}}} } A_{\alpha} \left( {t^{ \prime}}
\right)^{2}} \right\}} \approx
\]
\begin{equation}
\label{eq5}
 \approx {\frac{{\exp \left( {{\frac{{i}}{{\hbar}} }\vec {p}\;\vec {r}}
\right)}}{{\sqrt {V}}} }\exp {\left\{ { - {\frac{{i}}{{\hbar}}
}\varepsilon _{\vec {p}} t + {\frac{{ie_{0}}} {{c\hbar \omega}}
}{\sum\limits_{\alpha = 1}^{3} {{\frac{{p_{\alpha}  A_{\alpha}
^{\left( {0} \right)}}} {{m_{\alpha} }}}\sin \omega t}}}
\right\}}.
\end{equation}

In formula (\ref{eq5}), $V$ is the system volume, $\varepsilon
_{\vec {p}} = {\sum\limits_{\alpha} ^{3} {p_{\alpha} ^{2} /
2m_{\alpha}} }  $ is the energy of electron having momentum $\vec
{p}$. We have omitted the quadratic components in $A_{\alpha}
^{\left( {0} \right)} $ in the exponent, when obtaining formula
(\ref{eq5}). We shall find the electron wave function in the
presence of scattering centers by perturbation theory. This
function can be set in the following form:
\begin{equation}
\label{eq6}
\psi _{\vec {p}} = \psi _{\vec {p}}^{\left( {0} \right)} + \psi _{\vec
{p}}^{\left( {1} \right)} ,
\end{equation}
\noindent where $\psi _{\vec {p}}^{\left( {1} \right)} $ satisfies
the equation:
\begin{equation}
\label{eq7}
i\hbar {\frac{{\partial}} {{\partial t}}}\psi _{\vec {p}}^{\left( {1}
\right)} - \hat {H}^{\left( {0} \right)}\psi _{\vec {p}}^{\left( {1}
\right)} = {\sum\limits_{i = 1}^{N} {U\left( {\vec {r} - \vec {R}_{i}}
\right)\,\psi _{\vec {p}}^{\left( {0} \right)}}}  .
\end{equation}
We shall write the solution of Eq. (\ref{eq7}) as the expansion in
functions (\ref{eq5}):
\begin{equation}
\label{eq8} \psi _{\vec {p}}^{\left( {1} \right)} =
{\sum\limits_{\vec {p^{\prime}}} {C\left( {\vec {p},\,\vec
{p^{\prime}} ;\,t} \right)\,\,\psi _{\vec {p^{\prime}}}^{\left(
{0} \right)}}} .
\end{equation}
By substituting $\psi _{\vec {p}}^{\left( {1} \right)} $ in Eq.
(\ref{eq7}), multiplying both sides of that equation by
$\Psi_{\vec {p}}^{{({0})}^{\ast}} $, and integrating over $\vec
{r}$, we obtain

\[
i\hbar {\frac{{\partial}} {{\partial t}}}C\left( {\vec {p},\,\vec
{p^{\prime}} ;\,t} \right)=\frac{1}{V}\int   d \vec {r}\exp\Biggl(
\frac{i}{\hbar}(\vec {p}-\vec {p}^{\; \prime})\vec{r}
\Biggr)\times
\]
\[
\times\sum\limits_{j=1}^N U ( \vec{r}-\vec{R}_j )\exp \Biggl\{
-\frac{i}{\hbar} (\varepsilon_{\vec{p}} - \varepsilon_{\vec
{p^{\prime}}})t+
\]
\begin{equation}
+\frac{i e_0}{\hbar \omega c} \sum\limits_{\alpha = 1}^3
A_{\alpha}^{(0)} \Biggl(\frac{p_{\alpha}-
p_{\alpha}^{\prime}}{m_{\alpha}} \Biggr) \sin \omega t \Biggr\}.
\end{equation}

Integrating both sides of Eq. (9) between 0 and $t$ and taking into account the
identity
\begin{equation}
\label{eq9}
e^{ - i\lambda \sin \omega t} = {\sum\limits_{l = - \infty} ^{\infty}
{I_{l} \left( {\lambda}  \right)e^{ - il\omega t}}}
\end{equation}
(with $I_{l}$\textit{($\lambda )$} being the Bessel function), we
obtain:
\[
 C\left( {\vec {p},\,\vec {p^{\prime}} ;\,t} \right) = {\frac{{1}}{{i\hbar
V}}}\int d\vec {r}\exp \left( {{\frac{{i}}{{\hbar}} }\left( {\vec
{p} - \vec {p^{\prime}}}  \right)\vec {r}} \right)\times
\]
\[
\times\sum\limits_{j = 1}^{N} {U\left( {\vec {r} - \vec {R}_{j}}
\right)}\sum\limits_{l = - \infty} ^{\infty}  {I_{l} \left(
{\frac{{e_{0} }}{{\hbar \omega \,c}}}\sum\limits_{\alpha = 1}^{3}
{A_{\alpha} ^{\left( {0} \right)}\frac{{p_{\alpha}  - p_{\alpha}
^{ \prime}} } {{m_{\alpha}} } } \right)}\times
\]
\begin{equation}
\label{eq10} \times \frac{\exp \left\{ \left[  - \frac{i}{\hbar}
\left( \varepsilon _{\vec {p}} - \varepsilon_{\vec {p^{\prime} }}
\right) + il\omega \right]\,t \right\} - 1} {-
{\frac{{i}}{{\hbar}} }\left( {\varepsilon _{\vec {p}} -
\varepsilon _{\vec {p^{\prime}}} } \right) + il\omega}.
\end{equation}

Using Eq. (\ref{eq10}), one may find the probability for an
electron to pass from the state $\vec {p}$ into the state $\vec
{p}^{\; \prime} $ in a unit time, as a result of scattering by an
impurity in the field of an electromagnetic wave:
\begin{equation}
\label{eq11} P_{\vec {p},\vec {p^{\prime}}}  =
{\frac{{d}}{{dt}}}{\left| {C\left( {\vec {p},\,\vec {p^{\prime}}
;t} \right)} \right|}^{2}.
\end{equation}
Substituting relation (\ref{eq10}) in Eq. (\ref{eq11}), we obtain:
\[
 P_{\vec {p},\,\vec {p^{\prime}}}  = \frac{1}{\hbar ^{2}V^{2}}\Biggr|
\int d\vec {r}\exp \left( \frac{i}{\hbar} \left( \vec {p} - \vec
{p^{\prime}} \right)\,\vec {r} \right)\times
\]
\[
 \times \sum\limits_{j =
1}^{N} U\left( \vec {r} - \vec {R}_{j} \right)
\Biggr|^{2}\times\]
\[\times\Biggl\{ 2 \sum\limits_{l = - \infty} ^{\infty}
I_{l}^{2} \left( \frac{e_{0}} {\hbar \omega
\,c}\sum\limits_{\alpha = 1}^{3} A_{\alpha} ^{\left( {0} \right)}
\frac{\left( p_{\alpha}  - p_{\alpha }^{ \prime}
\right)}{m_{\alpha}}  \right)\times
\]
\begin{equation}
\label{eq12}
 \times\frac{{\sin \left( {\Omega - l\omega}
\right)t}}{{\Omega - l\omega}}  + {\sum\limits_{l \ne l^{ \prime}}
{\left( { \cdot \cdot \cdot}  \right)}}   \Biggr\}.
\end{equation}
In Eq. (\ref{eq12}), $\Omega \equiv {\frac{{1}}{{\hbar}} }\left(
{\varepsilon _{\vec {p}} - \varepsilon _{\vec {p^{\prime}}} }
\right).$

The terms with \textit{l $ \ne $ l}$^{ \prime} $ are not written
explicitly in Eq. (\ref{eq12}) since they do not contain resonance
multipliers and therefore disappear at $t \to \infty $. Passing to
the limit $t \to \infty $ in Eq. (\ref{eq12}) and taking into
account that ${\frac{{\sin xt}}{{x}}} \to \pi \delta \left( {x}
\right)$ in this case, we obtain the following expression for
$P_{\vec {p},\,\vec {p^{\prime}}} $:
\[
 P_{\vec {p},\,\vec {p^{\prime}}}  = {\frac{{2\pi}} {{\hbar V^{2}}}}{\left|
{\int {d\vec {r}\exp \left( {{\frac{{i}}{{\hbar}} }\left( {\vec
{p} - \vec {p^{\prime}}}  \right)\,r} \right){\sum\limits_{j =
1}^{N} {U\left( {\vec {r} - \vec {R}_{j}}  \right)}}} }
\right|}^{2}\times
\]
\[
 \times \sum\limits_{l = - \infty} ^{\infty}  I_{l}^{2} \left(
\frac{{e_{0}}} {\hbar \omega \,c}\sum\limits_{\alpha = 1}^{3}
{A_{\alpha}^{( 0)} } \frac{\left( {p_{\alpha}  - p_{\alpha }^{
\;\prime}} \right)} {m_{\alpha}}\right)\times
\]
\begin{equation}
\label{eq13} \times\delta \left( {\varepsilon _{\vec {p}} -
\varepsilon _{\vec {p^{\prime}}}  - l\hbar \omega}
\right).
\end{equation}

Expression (\ref{eq13}) depends explicitly on all the coordinates
of ions ${\left\{ {\vec {R}_{j}}  \right\}}$. That is why it
should be averaged over all the possible ion configurations.
Taking into account that
\[
 {\left| {\int {d\vec {r}\exp \left( {{\frac{{i}}{{\hbar}} }\left( {\vec {p}
- \vec {p}^{~\prime}}  \right)\,\vec {r}} \right){\sum\limits_{j =
1}^{N} {U\left( {\vec {r} - \vec {R}_{j}}  \right)}}} }
\right|}^{2} =
\]
\[
= {\sum\limits_{j = 1}^{N} {{\left| {\int {d\vec {r}\exp \left(
{{\frac{{i}}{{\hbar}} }\left( {\vec {p} - \vec {p^{\prime}}}
\right)\,\vec {r}} \right)U\left( {\vec {r} - \vec {R}_{j}}
\right)} ^{2}} \right|} +}}
\]
\[
 + {\sum\limits_{j \ne j^{ \prime}}  {\int {d\vec {r}_{1} \exp \left(
{{\frac{{i}}{{\hbar}} }\left( {\vec {p} - \vec {p^{\prime}}}
\right)\,\vec {r}_{1}} \right)\,}}}  U\left( {\vec {r}_{1} - \vec
{R}_{j}} \right)\times
\]
\begin{equation}
\label{eq14} \times \int d\vec {r}_{2} \exp \left( { -
{\frac{{i}}{{\hbar}} }\left( {\vec {p} - \vec {p^{\prime}}}
\right)\,\vec {r}_{2}} \right)\,U\left( {\vec {r}_{2} - \vec
{R}_{j^{ \prime}} } \right) ,
\end{equation}
\noindent one may show (e.g., see [2], p. 672) that the second
term in Eq. (\ref{eq14}) turns to zero, when averaging over the
positions of chaotically distributed scattering centers.
Therefore, the averaging results in
\[
\label{eq15} {\left\langle {{\left| {\int {d\vec {r}\exp \left(
{{\frac{{i}}{{\hbar }}}\left( {\vec {p} - \vec {p}^{~\prime}}
\right)\,\vec {r}} \right){\sum\limits_{j = 1}^{N} {U\left( {\vec
{r} - \vec {R}_{j}}  \right)} }}}  \right|}^{2}} \right\rangle}
=
\]
\begin{equation}
= N{\left| {\int {d\vec {r}\exp \left( {{\frac{{i}}{{\hbar}}
}\left( {\vec {p} - \vec {p^{\prime}}}  \right)\,\vec {r}}
\right)U\left( {\vec {r}} \right)}}  \right|}^{2}
\end{equation}.
In Eq. (16), $N$ is the number of ions in volume $V$, i.e.
\begin{equation}
{N=Vn}_{a},
\end{equation}
\noindent
with $n_{a}$ being the ion concentration.

Taking into account the explicit expression for $U\left( {\vec
{r}} \right)$ according to Eq. (\ref{eq2}), integral (16) can be
easily calculated:
\[
\label{eq16} \int d\vec {r}\exp \left( \frac{i}{\hbar}\left( \vec
{p} - \vec {p^{\prime}}  \right)\vec {r} \right)U\left( \vec {r}
\right)
=
\]
\begin{equation}
=  {\frac{{4\pi e_{0}^{2}}} {{\varepsilon _{0}}} }{\left\{ {\left(
{{\frac{{\vec {p} - \vec {p^{\prime}}} }{{\hbar}} }} \right)^{2} +
{\frac{{1}}{{r^{2}_{\mathrm{D}}}}}} \right\}}^{ - 2}.
\end{equation}

The average of $P_{\vec {p},\,\vec {p}^{\; \prime}}  $ over all
the ion configurations can be obtained from Eqs. (14) and (18) as
\[
 {\left\langle {P{}_{\vec {p},\vec {p^{\prime}}} ^{}}  \right\rangle}  =
{\frac{{\left( {2\pi \,\hbar}
\right)^{3}}}{{V}}}{\frac{{4e_{0}^{4} }}{{\varepsilon _{0}^{2}}}
}{\frac{{n_{a}}} {{{\left\{ {\left( {\vec {p} - \vec {p^{\prime}}}
\right)^{2} + {\frac{{\hbar ^{2}}}{{r_{\rm D}^{2}}} }}
\right\}}^{2}}}}\times
\]
\[
 \times {\sum\limits_{l = - \infty} ^{\infty}  {I_{l}^{2} \left(
{{\frac{{e_{0}}} {{\hbar \,\omega c}}}{\sum\limits_{\alpha =
1}^{3} {A_{\alpha} ^{\left( {0} \right)} {\frac{{\left(
{p_{\alpha}  - p^{ \prime}_{\alpha}}   \right)}}{{m_{\alpha}} }
}}}} \right)\delta \left( {\varepsilon _{\vec {p}} - \varepsilon
_{\vec {p^{\prime}}}  - l\hbar \omega} \right)}} .
\]
\begin{equation}
\label{eq17}
\end{equation}
The integral of collisions of electrons with ions in the presence
of electromagnetic field looks as follows:
\begin{equation}
\label{eq18} \left( {{\frac{{\partial f}}{{\partial t}}}}
\right)_{\mathrm{coll}} = - \sum \limits_{\vec {p^{\prime}}}
{{\left\langle {P_{\vec {p}\vec {p^{\prime}}} } \right\rangle}
f\left( {\vec {p}} \right) + {\sum\limits_{\vec {p^{\prime}}}
{{\left\langle {P_{\vec {p}_{1} \vec {p^{\prime}}} }
\right\rangle}} } f\left( {\vec {p^{\prime}}} \right)}.
\end{equation}

In Eq. (\ref{eq18}), $f\left( {\vec {p}} \right)$ is the
distribution function of electrons over their momentums $\vec
{p}$. Since we consider the multivalley semiconductors, the
distribution function can be different in different valleys. We
further write $f^{\left( {i} \right)}\left( {\vec {p}} \right)$,
meaning the distribution function in the $i$th valley.

\section{Light Absorption under Non-isotropic Impurity Scattering}

We substitute now Eq. (\ref{eq17}) in Eq. (\ref{eq18}) and proceed
from the summation over $\vec {p^{\prime}} $ to the integration.
This results in the following form of the collision integral for
the $i$th valley:
\[
 \left( {{\frac{{\partial f^{\left( {i} \right)}}}{{\partial t}}}}
\right)_{\mathrm{coll}} =
\]
\[
= {\frac{{4e_{0}^{4}}} {{\varepsilon _{0}^{2}}} }n_{a}
{\sum\limits_{l = - \infty} ^{\infty}  {\int {d\;\vec
{p^{\prime}}{\frac{{f^{\left( {i} \right)}\left(
{\vec{p^{\prime}}} \right) - f^{\left( {i} \right)}\left( {\vec
{p}} \right)}}{{{\left\{ {\left( {\vec {p} - \vec{p^{\prime}}}
\right)^{2} + \left( {\hbar / r_{\rm D}} \right)^{2}}
\right\}}^{2}}}}\times}} }
\]
\begin{equation}
\label{eq19}
 \times I_{l}^{2} \left( {{\frac{{e_{0}}} {{\hbar \omega
\,c}}}{\sum\limits_{\alpha = 1}^{3} {A_{\alpha} ^{\left( {0}
\right)}} }{\frac{{p_{\alpha}  - p_{\alpha}  ^{  \prime}}
}{{m_{\alpha}} } }} \right)\delta \left( {\varepsilon _{\vec {p}}
- \varepsilon _{\vec{p^{\prime}}}  - l\hbar \omega}  \right).
\end{equation}

We assume that the distribution function $f_{\left( {\vec {p}}
\right)}^{\left( {i} \right)} $ is normalized to the concentration
in the $i$th valley $n_{i}$:
\begin{equation}
\label{eq20} \int {d\vec {p}f^{\left( {i} \right)}\left( {\vec
{p}} \right) = n_{i}}.
\end{equation}

In thermal equilibrium, all $n_{i} $ are the same. Under heating
of electrons or a unidirectional pressure applied, the filling of
different valleys can be also different.

We obtain the energy absorbed by an electron of the $i$th valley
in a unit time from Eq. (\ref{eq19}) after multiplying it by
$\varepsilon _{\vec {p}} $ and integrating over
$\mathord{\buildrel{\lower3pt\hbox{$\scriptscriptstyle\rightharpoonup$}}\over
{p}} $. At this, if we make substitutions
$\mathord{\buildrel{\lower3pt\hbox{$\scriptscriptstyle\rightharpoonup$}}\over
{p}} \Leftrightarrow \vec{p^{\prime}} $ and $l \Leftrightarrow -
l$ in the term which is proportional to $f^{\left( {i}
\right)}\left( {\vec{p^{\prime}}} \right)$ as well as define
$\varepsilon _{\vec{p^{\prime}}}  $ through $\varepsilon _{\vec
{p}} $ using the \textit{$\delta $}-function, we obtain
\[
 P^{\left( {i} \right)} \equiv \int {d\vec {p}\varepsilon _{\vec {p}}}
\left( {{\frac{{\partial f^{\left( {i} \right)}}}{{\partial t}}}}
\right)_{st} =
\]
\[
= - {\frac{{4e_{0}^{4}}} {{\varepsilon _{0}^{2}}} }n_{\alpha}
{\sum\limits_{l = - \infty} ^{\infty}  {\hbar \omega l\int
{{\frac{{d\vec {p}d\vec{p^{\prime}} f^{\left( {i} \right)}\left(
{\vec {p}} \right)}}{{{\left\{ {\left( {\vec {p} - \vec
{p^{\prime}}} \right)^{2} + \left( {\hbar / r_{\rm D}}
\right)^{2}} \right\}}^{2}}}}} \times}}
\]
\begin{equation}
\label{eq21}
 \times I_{l}^{2} \left( {{\frac{{e_{0}}} {{\hbar \omega
c}}}{\sum\limits_{\alpha = 1}^{3} {A_{\alpha} ^{\left( {0}
\right)}} }{\frac{{p_{\alpha}  - p_{\alpha} ^{ \prime}} }
{{m_{\alpha}} } }} \right)\delta \left( {\varepsilon _{\vec {p}} -
\varepsilon _{\vec {p^{\prime}}}  - l\hbar \omega}  \right).
\end{equation}

From now on, we consider only one-phonon transitions, i.e. $l=\pm $1.

We have in this approximation:
\begin{equation}
\label{eq22}
P^{i} = P^{\left( {i} \right)}\left( { +}  \right) - P^{\left( {i}
\right)}\left( { -}  \right),
\end{equation}
\noindent where
\[
 P^{\left( {i} \right)}\left( {\pm}  \right) = \pm {\frac{{4e_{0}^{4}
}}{{\varepsilon _{0}^{2}}} }n\hbar \omega \int {{\frac{{d\vec
{p}d\vec{p^{\prime}} f^{\left( {i} \right)}\left( {\vec {p}}
\right)}}{{{\left\{ {\left( {\vec {p} - \vec{p^{\prime}}}
\right)^{2} + \left( {\hbar / r_{\rm D}^{}}  \right)^{2}}
\right\}}^{2}}}}} \times
\]
\begin{equation}
\label{eq23}
 \times I_{1}^{2} \left( {{\frac{{e_{0}}} {{\hbar \omega
c}}}{\sum\limits_{\alpha = 1}^{3} {A_{\alpha} ^{\left( {0}
\right)} {\frac{{p_{\alpha}  - p_{\alpha} ^{ \prime}} }
{{m_{\alpha}} } }}}} \right)\delta \left( {\varepsilon _{\vec {p}}
- \varepsilon _{\vec {p^{\prime}}} \pm \hbar \omega}  \right).
\end{equation}

The sign (+) means an increase of the electron system energy
(i.e., absorption), while the sign (--) means a decrease of this
energy (i.e., emission).

As the estimations made for all frequencies of the optical range
show, the argument of  the function $I_{1} ( \ldots)$ in Eq.
(\ref{eq23}) is far below unity. That is why we can consider only
the first term of the Taylor series of $I_{1}({\ldots} )$ in Eq.
(\ref{eq23}). We then have
\[
P^{( {i})}\left( {\pm}  \right) \cong \pm \frac{e_{0}^{6}
n_{\alpha}}  {\varepsilon _{0}^{2} c^{2}\hbar \omega} \int \frac
{d\vec {p}d\vec{p^{\prime}} f^{( {i})}( {\vec {p}})\delta \biggl\{
{\varepsilon _{\vec {p}} - \varepsilon _{\vec {p}^{ \; \prime}}
\pm \hbar \omega} \biggr \}}{ \biggl\{ ( {\vec {p} - \vec
{p^{\prime}}})^{2} + ( {\hbar / r_{\rm D}} )^{2}
\biggr\}^{2}}\times
\]
\begin{equation}
\label{eq24} \times\left( {{\sum\limits_{\alpha = 1}^{3}
{A_{\alpha} ^{0}}} {\frac{{p_{\alpha}  - p_{\alpha}  ^{ \prime}}
}{{m_{\alpha}} } }} \right)^{2}.
\end{equation}

The distribution function must be specified to calculate integral
(\ref{eq24}). To be able to analyze further the general case, we
assume that concentrations ($n_{i})$ and temperatures
(\textit{$\theta $}$_{i})$ in different valleys can be different.

We assume that
\begin{equation}
\label{eq25} f^{\left( {i} \right)}\left( {\vec {p}} \right) =
{\frac{{n_{i}}} {{\left( {2\pi \theta _{i}}  \right)^{3 / 2}m_{
\bot}  \sqrt {m_{\|}} } } }\exp \left( { - {\frac{{\varepsilon
_{\vec {p}}}} {{\theta _{i}}} }} \right).
\end{equation}

In the principal axes of the mass tensor,
\begin{equation}
\label{eq26} \varepsilon _{\vec {p}} = {\frac{{p_{ \bot} ^{2}}}
{{2m_{ \bot}} } } + {\frac{{p_{ \|} ^{2}}} {{2m_{ \|}} } }.
\end{equation}

In addition,
\begin{equation}
 {\sum\limits_{\alpha = 1}^{3} {A_{\alpha} ^{\left( {0}
\right)} {\frac{{p_{\alpha}  - p_{\alpha} ^{ \prime}} }
{{m_{\alpha}} } } = {\frac{{\hbar \gamma}} {{m_{ \bot}} } },}}
\end{equation}
\noindent
where
\[
 \gamma \equiv \vec {A}_{ \bot} ^{\left( {0} \right)} \vec {q}_{ \bot}  +
{\frac{{m_{ \bot}} } {{m_{ \|}} } }A_{ \|} ^{\left( {0} \right)}
q_{ \|} =
\]
\[
 = \vec {A}^{\left( {0} \right)}\vec {q} + \left(
{{\frac{{m_{ \bot}} } {{m_{ \|}} } } - 1} \right)\left( {\vec
{A}^{\left( {0} \right)}\vec {q}} \right)\left( {\vec {i}_{0} \vec
{q}} \right),
\]
\begin{equation}
\label{eq27}
 \hbar \vec {q} \equiv \vec {p} - \vec {{p}^{\prime}}.
\end{equation}

In Eq. (\ref{eq27}), $\vec {i}_{0} $ is the ort which specifies
the direction of the line of rotation of the mass ellipsoid. The
direction of this ort coincides with that of the position of the
$i$th valley in the laboratory system.

The angular dependence of the energy makes the integration much more
difficult. That is why it is convenient to use a deformed coordinate system,
in which the surfaces of equal energy are spherical.

We introduce new variables therefore:
\[
 p_{ \bot} ^{\ast}  = p_{ \bot}  ,\,p_{ \|} ^{\ast} =
\left( {{\frac{{m_{ \bot}} } {{m_{ \|}} } }} \right)^{1 / 2}p_{\|}
;\;\;q_{ \bot} ^{\ast}  = q_{ \bot}  ;
\]
\begin{equation}
\label{eq28}
 q_{ \|} ^{\ast}  = \left( {{\frac{{m_{ \bot}} }
{{m_{ \|} }}}} \right)^{1 / 2}q_{ \|}  .
\end{equation}

At this,
\begin{equation}
\varepsilon _{\vec {p}} = \left( {p^{\ast}}  \right)^{2} / 2m_{
\bot}  .
\end{equation}
Expression (\ref{eq24}) acquires the following form in new variables:
\[
 P_{( {\pm})}^{( {i})} = \pm
\frac{e_{0}^{6} n_{a} }{\varepsilon_{0}^{2} c^{2}\omega}
\frac{m_{\|} } {m_{ \bot} }\times
\]
\[
\times \int \frac{d\vec {p}^{~\ast} d\vec {q}^{~\ast} f^{( i )}(
\varepsilon _{p^{\ast}} )\delta \left\{ \frac{( \hbar
q^{\ast})^{2}}{2m_{ \bot}}  - \frac{\hbar} {m_{ \bot} } p^{\ast}
q^{ *} \cos \nu ^{ *} \pm \hbar \omega \right\}\gamma
^{2}}{\left\{ q_{ \bot}^{ *2} + \frac{m_{ \|} } {m_{ \bot}}
 q_{ \|}^{ *2} +( {1 / r_{\rm D} })^{2}
\right\}^{2}}.
\]
\begin{equation}
\label{eq29}
\end{equation}

Now we take into account that $d\vec {p}^{~*}  \to p^{\ast
2}dp^{\ast }d\Omega _{p^{\ast}}  \to p^{ *2}dp^{\ast} \sin \nu ^{
*} d\nu ^{ * }d\varphi ^{ *} $.

The integral over \textit{$\varphi $*} can be easily calculated
(since nothing depends on \textit{$\varphi $}*). The integral over
\textit{$\nu^{\ast}$} can be calculated using the $\delta
$-function:
\[
{\int\limits_{0}^{\pi}  {d\nu ^{ *} \sin \nu ^{ *} \delta {\left\{
{{\frac{{\left( {\hbar q^{ *}}  \right)^{2}}}{{2m_{ \bot}} } } -
{\frac{{\hbar}} {{m_{ \bot}} } }p^{ *} q^{ *} \cos \nu ^{ *} \pm
\hbar \omega}  \right\}}}}  = {\frac{{m_{ \bot}} } {{\hbar p^{ *}
q^{ *}} }}.
\]
\begin{equation}
\label{eq30}
\end{equation}

Equality (\ref{eq30}) is real under the condition that
\begin{equation}
\label{eq31} {\left| {\cos \nu ^{ *}}  \right|} = {\left| {\pm
\hbar \omega - {\frac{{\left( {\hbar q^{ *}}  \right)^{2}}}{{2m_{
\bot}} } }} \right|} \Biggl/ {\frac{{\hbar q^{ *} p^{ *}} }{{m_{
\bot}} } } \le 1.
\end{equation}

Condition (\ref{eq31}) means that the argument of the
\textit{$\delta $}-function at a specified $q$* can be equal to
zero. In other words, inequality (\ref{eq31}) determines the
limits of the integration over $q$*. We find from Eq.
(\ref{eq31}):
\[
 \hbar q_{\max}  \left( {\pm}  \right) = p^{ *}  + \sqrt {p^{ *2}\pm
2m_{ \bot}  \hbar \omega},
\]
\begin{equation}
\label{eq32}
 \hbar q_{\min}  \left( {\pm}  \right) = {\left| { - p^{ *}  + \sqrt {p^{ *2}\pm 2m_{ \bot}  \hbar \omega}}
 \right|}.
\end{equation}

After the integration over the angles of the vector $\vec {p}^{~*}
$, we find from Eq. (\ref{eq29}):
\[
 P_{}^{\left( {i} \right)} ( + ) = {\frac{{e_{0}^{6} n_{a} n_{i} \sqrt {m_{\|}} } } {{\sqrt {2\pi}  \,\theta _{i}^{3 / 2} \varepsilon
_{0}^{2} c^{2}\hbar \omega}} }{\int\limits_{0}^{\infty}
{d\varepsilon e^{ - \varepsilon / \theta _{i}}} }
{\int\limits_{q_{\min} ^{\left( { +}  \right)} }^{q_{\max}
^{\left( { +}  \right)}}  {dq^{ *} q^{ *}} } \times
\]
\begin{equation}
\label{eq33}
 \times \int {{\frac{{d\Omega _{q^{~*}}  \gamma ^{2}\left( {\vec {q}^{ *}}
\right)}}{{{\left\{ {q^{ *2} + {\frac{{m_{ \|}} } {{m_{ \bot}} }
}q_{ \|} ^{ *2} + \left( {1 / r_{\rm D}}  \right)^{2}}
\right\}}^{2}}}}}.
\end{equation}

Unlike Eq. (\ref{eq33}), the integral over \textit{$\varepsilon $}
for the $P^{(i)}(-)$ quantity must be calculated within limits
from $\hbar \omega $ to $\infty $ (since only electrons with
energy $\varepsilon \ge \hbar \omega $ can emit $\hbar \omega $
quanta). If we make a shift $\varepsilon \to \varepsilon - \hbar
\omega $ in the expression for $P^{(i)}(-)$, we obtain that
\[
P^{\left( {i} \right)}( - ) = - \exp \left( { - {\frac{{\hbar \omega
}}{{\theta _{i}}} }} \right)P^{\left( {i} \right)}\left( { +}  \right).
\]

The latter integral over the angles of the vector $\vec {q}^{ *} $
in Eq. (\ref{eq33}) can be easily calculated with a result
\[
 y\left( {q^{ *}}  \right) \equiv \int {{\frac{{d\Omega _{q^{*}}  \gamma
^{2}\left( {\vec {q}^{~*}}  \right)}}{{{\left\{ {q^{ *2} +
{\frac{{m_{ \|}} } {{m_{ \bot}} } }q_{ \|} ^{*2} + \left( {1 /
r_{\rm D}} \right)^{2}} \right\}}^{2}}}}} =
\]
\[
 = {\frac{{\pi}} {{q^{ *2}}}}{\left\{ {\left( {A_{ \bot} ^{\left( {0}
\right)}}  \right)^{2}B_{1} \left( {q^{ *}}  \right) + 2\left(
{A_{\|}^{\left( {0} \right)}}  \right)^{2}{\frac{{m_{ \bot}} }
{{m_{ \|} }}}B_{2} \left( {q^{ *}}  \right)} \right\}}\times
\]
\begin{equation}
\label{eq34}
 \times\left( {{\frac{{m_{ \bot} }}{{m_{ \|}  - m_{
\bot}} } }} \right)^{2}.
\end{equation}

We have made the following designations in Eq. (\ref{eq34}):
\[
 B_{1} \left( {q^{ *}}  \right) = {\frac{{1}}{{b^{2}}}} + {\frac{{1 -
b^{2}}}{{b^{3}}}}{\rm arctg}{\frac{{1}}{{b}}},
\]
\[
 B_{2} \left( {q^{ *}}  \right) = - {\frac{{1}}{{1 + b^{2}}}} +
{\frac{{1}}{{b}}}{\rm arctg}{\frac{{1}}{{b}}},
\]
\begin{equation}
\label{eq35}
 b^{2} = {\frac{{m_{ \bot}} } {{m_{ \|}  - m_{ \bot}} } }\left( {1 + 1 /
\left( {q^{ *} r_{\rm D}}  \right)^{2}} \right).
\end{equation}

The double integral  in Eq. (\ref{eq33}) can be reduced to a
single one by integration by parts:
\[
 \int\limits_{0}^{\infty}  d\varepsilon e^{ - \varepsilon /
\theta _{i} } \int\limits_{q_{\min}^{(-)}}^{{q_{\max}^{( +)}}}
dq^{*} q^{*} y( q^{*}) =
\]
\[
=\theta _{i} \int\limits_{0}^{\infty} d\varepsilon e^{ -
\varepsilon / \theta _{i}} \Biggl\{ [ q^* y(q)]_{q ={
q_{\max}(+)}}\frac{dq_{\max} (  + )}{d\varepsilon}-
\]
\begin{equation}
\label{eq36} - [ q^* y( q)]_{q = q_{\min}(+)} \frac{dq_{\min}
(+)}{d\varepsilon} \Biggr\}.
\end{equation}

We substitute now Eqs. (\ref{eq36}) and (\ref{eq34}) in Eq.
(\ref{eq33}) and introduce a dimensionless variable
\textit{ї=$\varepsilon $/$\theta $}$_{i}$. In addition, we rewrite
the quantities $A_{ \bot} ^{\left( {0} \right)} \,\,i\,\,A_{ \|}
^{\left( {0} \right)} $ in the laboratory coordinate system, which
are written in Eq. (\ref{eq38}) in principal axes of the $i$th
ellipsoid (valley). This means that
\[
\label{eq37}
 ( A_{ \|} ^{( {0})})^{2} = \left( {\vec {i}_{0} \vec
{q}_{0}}  \right)^{2} A^{{( {0})^{2}}},
\]
\begin{equation}
( A_{ \bot }^{( {0} )} )^{2} = A^{{( {0} )}^{2}} -( {\vec {i}_{0}
\vec {q}_{0}})^{2} A^{{( {0})^{2}}}.
\end{equation}

In Eq. (41), $\vec {q}_{0} $ is an ort, which characterizes wave
polarization. The ort $\vec {i}_{0} $ defines the direction of the
$i$th valley position.

We obtain the following expression as a result of the operations
made above:
\[
P^{( {i} )}( { +} ) = {\frac{e_{0}^{6} n_{a} n_{i}} {4\varepsilon
_{0}^{2} c^{2}\hbar \omega} }\Biggl( \frac{2\pi m_{ \|}} {\theta
_{i}} \Biggr)^{1 / 2}\frac{A^{{( {0} )}^{2}}}{( m_{ \|} - m_{
\bot} )^{2}}\times
\]
\begin{equation}
\label{eq38} \times {\int\limits_{0}^{\infty}  {{\frac{{dxe^{ -
x}{\left[ {\Psi _{i} \left( {q_{\max}  \left( { +}  \right)}
\right) + \Psi _{i} \left( {q_{\min}  \left( { +}  \right)}
\right)} \right]}}}{{\sqrt {x\left( {x + \hbar \omega / \theta
_{i}} \right)}}} }}}.
\end{equation}

We have introduced the following designation in Eq. (\ref{eq38}):
\[
\Psi _{i} \left( {q^{ *}}  \right) = B_{1} \left( {q^{ *}}
\right) + \left( {\vec {i}_{0} \vec {q}_{0}}  \right)^{2}{\left[ {
- B{}_{1}\left( {q^{ *}} \right) + 2{\frac{{m_{ \bot}} } {{m_{
\|}} } }B_{2} \left( {q^{ *}} \right)} \right]}.
\]
\begin{equation}
\label{eq39}
\end{equation}
In dimensionless variables in accordance with Eq. (\ref{eq32}),
the quantities $q_{\max}(+)$ and $q_{\min}(+)$ in Eq. (\ref{eq38})
become as follows:
\[
 q_{\max}  \left( { +}  \right) = {\frac{{\left( {2m_{ \bot}  \theta _{i}}
\right)^{1 / 2}}}{{\hbar}} }{\left[ {x{}^{1 / 2} + \left( {x +
{\frac{{\hbar \omega}} {{\theta _{i}^{}}} }} \right)^{1 / 2}}
\right]},
\]
\begin{equation}
\label{eq40}
 q_{\min}  \left( { +}  \right) = {\frac{{\left( {2m_{ \bot}  \theta _{i}}
\right)^{1 / 2}}}{{\hbar}} }{\left[ { - x{}^{1 / 2} + \left( {x +
{\frac{{\hbar \omega}} {{\theta _{i}^{}}} }} \right)^{1 / 2}}
\right]}.
\end{equation}

\section{Absorption Coefficient}

We have found above the energy absorbed or emitted in a unit time.
Experimentally, the adsorption coefficient is measured, which
looks as follows:
\[
 K = \frac{\sum\limits_{i} ( P^{(i)}(+) +
P^{(i)}( - ) )}{\Pi } =
\]
\begin{equation}
\label{eq41}
 = \frac{\sum\limits_{i} (1 - \exp ( -
\frac{{\hbar \omega}} {\theta _{i} })){\rm P}^{(i)}( + ) }\Pi .
\end{equation}

In Eq. (\ref{eq41}), $\Pi $ is an electromagnetic flow, which impinges on a
semiconductor:
\begin{equation}
\label{eq42}
\Pi = {\frac{{\varepsilon _{0}^{1 / 2}}} {{8\pi}} }{\frac{{\omega
^{2}}}{{c}}}A^{(0)^{2}}.
\end{equation}

Substituting Eqs. (\ref{eq42}) and (\ref{eq38}) in Eq.
(\ref{eq41}), we obtain
\[
 K = \frac{(2\pi )^{3 / 2}e_{0}^{6} n_{a} m_{ \|} ^{1 /
2} }{\varepsilon _{0}^{5 / 2} c(m_{ \|}  - m_{ \bot} )^{2}\hbar
\omega ^{3}} \sum\limits_{i} \frac{n_{i}} {\sqrt { \theta _{i}}}
\Biggl( 1 - \exp \Biggl( - \frac{\hbar \omega} {\theta _{i}}
\Biggr)\Biggr)\times
 \]
 \begin{equation}
\label{eq43}
 \times \int\limits_{0}^{\infty} \frac{ dxe^{ -
x}\left\{ \Psi _{i} (q_{\max}  ( + ) + \Psi _{i} (q_{\min } ( + )
\right\}}{\sqrt {x(x + \hbar \omega / \theta _{i} )}} .
\end{equation}

Expression (\ref{eq43}) gives the general value of the adsorption coefficient under
anisotropic impurity scattering in multivalley semiconductors. Different
values of the filling of valleys ($n_{i} )$ can be connected with different
electron temperatures ($\theta _{i} )$ in valleys, or can be caused by the
unidirectional pressure (which shifts the valleys). In a state of
thermodynamic equilibrium and without unidirectional pressure, all the
$n_{i}$ and \textit{$\theta $}$_{i}$ values are identical. The temperatures $\theta _{i} $ can
become different when electrons are heated by an external electric field.
They can also become different under absorption of the polarized light with
sufficient intensity. That is, the polarization dependence of absorption can
appear in multivalley semiconductors at sufficiently high intensities. This
problem was theoretically studied in [3]. When $n_{i} $ and $\theta _{i} $
differ in different valleys, the balance equations of concentrations and
energy should be used to find them (e.g., see [4]).

The general expression for the absorption coefficient (\ref{eq43}) under impurity
scattering can be substantially simplified in the classical $\left( {\hbar
\omega / \theta _{i} < < 1} \right)$ and quantum $\left( {\hbar \omega /
\theta _{i} > > 1} \right)$ cases. Therefore, we shall analyze both these
cases.

R e g i o n~ o f~ c l a s s i c a l~ a b s o r p t i o n $\left(
{\hbar \omega / \theta _{i} < < 1} \right)$. According to Eq.
(\ref{eq40}), we have
\begin{equation}
\label{eq44} q_{\max}  \left( { +}  \right) \approx \left( {2m_{
\bot}  \theta _{i}} \right)^{1 / 2}{\frac{{2\sqrt {x}}} {{\hbar}}
},\,\,q_{\min}  \left( { +} \right) \approx 0.
\end{equation}

In Eq. (\ref{eq40}), the integration occurs over the dimensionless energy $x =
\varepsilon / \theta _{i} $. Then we take into account that the quantity
\begin{equation}
\label{eq45} b^{2} = {\frac{{m_{ \bot}} } {{m_{ \|}  - m_{ \bot}}
} }(1 + {\frac{{1}}{{(q^{ *} r_{\rm D} )^{2}}}}) \equiv b_{0}
^{2}(1 + {\frac{{1}}{{(q^{ *} r_{\rm D} )^{2}}}})
\end{equation}
\noindent has a very weak dependence on $x$, since (according to
estimations) $\left( {\vec {q}^{ *} r_{\rm D}}  \right)^{2} > >
1$. Here, $q^{ *} $ is taken at average energies (i.e., at $x
\approx 1)$.

We can see from Eqs. (\ref{eq39}) and (\ref{eq35}) that $\Psi _{i}
(q^{ *} )$ can depend on $x$ only due to the dependence of $b^{2}$
on $x$. One may easily see that $b^{2} \to \infty $ in this case
at $q^{ *}  = q_{\min}  ( + ) \approx 0$. According to Eqs.
(\ref{eq35}) and (\ref{eq39}), we obtain $\Psi _{i} (q_{\min}  ( +
)) \approx 0$. We now pay attention to the $\Psi _{i} (q_{\max}  (
+ ))$ value. At $q^{ *}  = q_{\max}  ( + )$, the value of $b^{2}$
is almost independent of $x$ since $(q_{\max}  ( + )r_{\rm D}
)^{2} >
> 1$ for all $x\sim $1. The dependence on $x$ becomes significant
only at small $x$, for which $\left( {q_{\max}  ( + )r_{\rm D}}
\right) \le 1^{}$. We shall find the value of $x = x_{\min}  $
from the condition $q{}_{\max} ( + )r_{\rm D} = 1$. We obtain from
Eq. (\ref{eq44}):
\begin{equation}
\label{eq46} x_{\min}  = {\frac{{1}}{{8}}}{\frac{{\hbar
^{2}}}{{m_{ \bot}  \theta _{i} r_{\rm D}^{2}}} }.
\end{equation}

Because of the stated above, when integrating approximately in Eq.
(\ref{eq38}), we can take out $\Psi _{i} $ of the integration sign
and truncate the integration over $x$ at $x = x_{\min}  $.
Therefore, we obtain at $\hbar \omega / \theta _{i} < < 1$:
\begin{equation}
\label{eq47}
 \int\limits_{0}^{\infty} \frac{dxe^{ - x}\Psi
_{i} {(q_{\max}  ( + ))}}{\sqrt {x(x + \hbar \omega / \theta _{i}
)}} \approx \Psi _{i} (\infty ){\int \limits_{x_{\min}}^{\infty}
\frac{dx}{x}e^{ - x}}.
\end{equation}
In Eq. (\ref{eq47}), we have set $q_{\max}  ( + ) \approx \infty
$. This is justified at $q_{\max}  ( + )r_{\rm D} > > 1$, as can
be seen from Eq. (\ref{eq45}). As will be seen below,
approximation (\ref{eq47}) meets the known logarithmic
approximation in the description of impurity scattering (so-called
Conwell---Weisskopf approximation).

According to Eqs. (\ref{eq39}) and (\ref{eq35}), we have:
\[
\Psi _{i} \left( {\infty}  \right) = {\frac{{1}}{{b_{0}^{3}}}
}{\left[ {b_{0} + \left( {1 - b_{0}^{2}} \right){\rm
arctg}{\frac{{1}}{{b_{0}}} }} \right]}\sin ^{2}\varphi _{i} +
\]
\begin{equation}
\label{eq48} + 2{\frac{{m_{ \bot}} } {{m_{ \|}} } }{\left[ { -
{\frac{{1}}{{1 + b_{0}^{2}}} } + {\frac{{1}}{{b_{0} }}}{\rm
arctg}{\frac{{1}}{{b_{0}}} }} \right]}\cos ^{2}\varphi _{i}
\end{equation}
\noindent
 with $\cos \varphi _{i} \equiv \vec {i}_{0} \vec {q}_{0}
$, i. e., $\varphi _{i} $ is an angle between the line of rotation
of the mass ellipsoid of the $i$th valley and the ort of the wave
polarization $ {\vec {q}_{0}} $.

Now we obtain a simpler form of the absorption coefficient for the
classical region from the general expression (41), using
approximation (\ref{eq47}):
\begin{equation}
\label{eq49} { K} = {\frac{{3\pi ^{3 /
2}}}{{2}}}{\frac{{e_{0}^{2}}} {{\varepsilon _{0}^{1 / 2}}}
}{\frac{{1}}{{c\omega ^{2}}}}{\sum\limits_{i} {n_{i} {\Biggl\{
{{\frac{{\sin ^{2}\varphi _{i}}} {{m_{ \bot}  \tau _{ \bot} (
{\theta _{i}}  )}}} + {\frac{{\cos ^{2}\varphi _{i}}} {{m_{ \|}
\tau _{\|} (\theta _{i} )}}}{ {} \Biggr\}}} }}} .
\end{equation}
In expression (\ref{eq49}), $\tau _{ \bot}  \left( {\theta _{i}}
\right)$ and $\tau _{ \|}  \left( {\theta _{i}}  \right)$ are the
respective components of the relaxation tensor under impurity
scattering. At this
\[
 \frac{1}{\tau _{ \bot} ( \theta _{i} )}
 = \frac{8}{3}\frac{e_{0}^{4}( 2m_{ \|} )^{1 / 2}}{\varepsilon
_{0}^{2} m_{ \bot}  \theta _{i}^{3 / 2}}\times
\]
\[
\times n_{a} \frac{b_{0}} {2} \Biggl[ b_{0} +( 1 - b_{0}^{2})\rm
{arctg}\frac{1}{b_{0}} \Biggr]\ln( C_{1} x_{\min})^{ - 1},
\]
\[
 \frac{1}{\tau _{  \|} ( \theta _{i} )} = {\frac{{8}}{{3}}}{\frac{{e_{0}^{4} \left( {2m_{\|}} \right)^{1 / 2}}}{{\varepsilon _{0}^{2} m_{ \|}  \theta
_{i}^{3 / 2}}} }\times
\]
\begin{equation}
\label{eq50}\times n_{a} b_{0} {\left[ { - b_{0} + \left( {1 +
b_{0}^{2}} \right){\rm arctg}{\frac{{1}}{{b_{0} }}}} \right]}\ln
\left( {C_{1} x_{\min}}   \right)^{ - 1}.
\end{equation}
 $\ln (C_{1} x_{\min}  )^{ - 1}$ appears in Eqs. (\ref{eq50}) because the integral on
the right part of Eq. (\ref{eq47}) is equal to

 \[
 {\int\limits_{x_{\min}}  ^{\infty}  {{\frac{{dx}}{{x}}}e^{ - x} = \ln \left(
{C_{1} x_{\min}}   \right)^{ - 1} - {\sum\limits_{k = 1}^{\infty}
{{\frac{{1}}{{kk}}}\left( { - x_{\min}}   \right)^{k}}}} } ,
\]
 where
$\ln C_{1} = 0.577...$ is the Euler constant. Since $x_{\min}  < <
1$, we have confined ourselves within only the logarithmic
approximation in Eq. (\ref{eq50}).

The components of the relaxation tensor under impurity scattering are
connected with the mobility tensor components by the following relations:
\begin{equation}
\label{eq51} \mu _{ \bot}  = {\frac{{8}}{{\sqrt {\pi}} }
}{\frac{{e_{0} \tau _{ \bot} \left( {\theta _{i}}  \right)}}{{m_{
\bot}} } }; \quad \mu _{ \|}  = {\frac{{8}}{{\sqrt {\pi}} }
}{\frac{{e_{0} \tau _{ \|} \left( {\theta _{i}} \right)}}{{m_{\|}}
} }.
\end{equation}

R e g i o n~ o f~ q u a n t u m~ a b s o r p t i o n $\left(
{\hbar \omega
>
> \theta _{i}}  \right)$.

In this case, we obtain from Eq. (\ref{eq40}):
\begin{equation}
\label{eq52}
q_{\max}  ( + ) \approx q_{\min}  ( + ) = \left( {{\frac{{2m_{ \bot}
}}{{\hbar}} }\omega}  \right)^{1 / 2} \equiv q_{\omega}  .
\end{equation}

Now integral (\ref{eq43}) can be easily estimated:
\[
 \int\limits_{0}^{\infty}  \frac{dxe^{ - x} \left\{\Psi_{i} (q_{\max}( + ) ) + \Psi_{i}
 (q_{\min}  ( + )) \right\} }{\sqrt {x(x +
\hbar \omega / \theta _{i} )}}\approx
\]
\begin{equation}
\approx 2\sqrt {\pi}  \Psi _{i} (q_{\omega} )\Biggl(\frac{\theta
_{i}} {\hbar \omega} \Biggr)^{1 / 2} .
\end{equation}
We may set $\Psi _{i} (q_{\omega}  ) \approx \Psi _{i} (\infty )$,
taking into account that $\left( {q_{\omega}  r_{\rm D}}  \right)
> > 1$. As a result, for the quantum region $\left( {\hbar \omega
> > \theta _{i}}  \right)$, we have from Eq. (41):
\begin{equation}
\label{eq53} { K} = \left( {{\frac{{2\pi}} {{\varepsilon _{0}}} }}
\right)^{5 / 2}{\frac{{e_{0}^{6} n_{a} m_{ \|} ^{1 / 2}}} {{c(m_{
\|}  - m_{ \bot} )^{2}\omega ^{2}}}}{\frac{{\sum {n_{i} \Psi _{i}
(\infty )}}} {{(\hbar \omega )^{3 / 2}}}}.
\end{equation}
The form of the function $\Psi _{i} (\infty )$ is specified by
formula (\ref{eq48}), where the explicit dependence on the
polarization angle $\varphi _{i} $ is given.

Thus, we have obtained a simple expression for the light adsorption
coefficient in the classical and quantum ranges of frequencies under the
dominating role of impurity (anisotropic) scattering.

To complete the picture, we also present the expression for the
adsorption coefficient in the case of the dominating role of
acoustic scattering. This will enable us to compare the
peculiarities of manifestation of various scattering mechanisms in
the phenomenon of light absorption by free carriers.\looseness=1

The absorption coefficient in the case of anisotropic acoustic
scattering was obtained in [1]. Before writing it down, we recall
(e.g., see [1]) that the components of the relaxation tensor as
functions of electron energy in semiconductors of the kind of
$n$-Ge and $n$-Si can be written in the following form under
acoustic scattering:\looseness=1
\[
\tau _{x} (\varepsilon ) = \tau _{y} (\varepsilon ) \equiv \tau _{ \bot}
(\varepsilon ) = \tau _{ \bot} ^{(0)} \left( {{\frac{{\theta _{i}
}}{{\varepsilon}} }} \right)^{1 / 2};
\]
\begin{equation}
\label{eq54} \tau _{ \|}  (\varepsilon ) = \tau _{ \|} ^{(0)}
\left( {{\frac{{\theta _{i}}} {{\varepsilon}} }} \right)^{1 / 2}.
\end{equation}

\noindent The general expression for the adsorption coefficient in
the whole frequency range under anisotropic acoustic scattering
has a following form [1]:
\[
{K} = - \frac{16\sqrt {\pi} } {3\sqrt {\varepsilon _{0}}
}\frac{e_{0}^{2}} {c\hbar}\sum\limits_{i} \frac{n_{i} \theta _{i}}
{\omega ^{3}}( 1 - e^{ - \hbar \omega / \theta _{i}})\times
\]
\begin{equation}
\label{eq55}
 \times\Biggl\{ \frac{\sin ^{2}\varphi _{i}} {m_{ \bot}
\tau _{ \bot}( \theta _{i}  )} + \frac{\cos ^{2}\varphi _{i}}
{m_{\|} \tau _{ \|}  (\theta _{i} )} \Biggr\}\Biggl\{
a_{i}^{3}\frac{d}{da_{i}} \Biggl( \frac{K_{1} (a_i )}{a_{i} }
\Biggr) \Biggr\} .
\end{equation}
[In Eq. (\ref{eq55}), we have corrected the misprints made in this
formula in [1]].

In Eq. (\ref{eq55}), $a_{i} = \hbar \omega / 2\theta _{i} $ and
${K}_{1} (a_{i} )$ is a Bessel function, whose asymptotic form is
as follows:
\begin{equation}
\label{eq56} {K}_{1} (x) = \left\{ \begin{array}{l}
 \frac{1}{x}\,\,\,\,\,\,\,\,\,\,\,\,\,\,\,\,\,\,\,\,\, \text{at}\,x \to 0, \\
 \sqrt {\frac{\pi} {2x}}\, e^{ - x}\, \, \text{at}\,\,\,x \to \infty .  \\
 \end{array} \right.
 \end{equation}
To avoid misunderstanding, we notice that $\tau _{ \bot ,\|}
^{(0)} $ in Eq. (\ref{eq54}) differs from $\tau _{ \bot , \|}
^{(0)} $ in [1] by a factor $\left( {{\frac{{\theta _{i}}}
{{\theta}} }} \right)^{1 / 2}$, since we have $\tau _{ \bot , \|}
(\theta ) = \left( {{\frac{{\theta _{i}}} {{\theta }}}} \right)^{1
/ 2}\tau _{ \bot , \|} ^{(0)} $ from Eq. (\ref{eq54}), while $\tau
_{ \bot , \|}  (\theta ) = \tau _{ \bot , \|} ^{(0)} $ in [1]. We
do not introduce the lattice temperature $\theta $ at all in this
paper. The electron temperatures $\theta _{i} $ stand everywhere,
which can coincide with the lattice temperature or differ from it.

We obtain the absorption coefficient in the classical and quantum ranges of
frequencies from the general formula (\ref{eq55}), using asymptotics (\ref{eq56}).

Therefore, we get in the classical range $(\hbar \omega < < \theta _{i} )$:
\[
 {K} = {\frac{{32\sqrt {\pi}} }
{{3}}}{\frac{{e_{0}^{2}}} {{\sqrt {\varepsilon _{0}}} }
}{\frac{{1}}{{c\omega ^{2}}}}{\sum\limits_{i} {n_{i} {\left\{
{{\frac{{\sin ^{2}\varphi _{i}}} {{m_{ \bot}  \tau _{ \bot} \left(
{\theta _{i}}  \right)}}} + {\frac{{\cos ^{2}\varphi _{i}}} {{m_{
\|} \tau _{ \|}  (\theta _{i} )}}}} \right\}}}} .
\]
\begin{equation}
\label{eq57}
\end{equation}
In the quantum range $(\hbar \omega > > \theta _{i} )$, we have,
respectively,
\[
{K} = {\frac{{4\pi}} {{3}}}{\frac{{e_{0}^{2}}} {{\sqrt
{\varepsilon _{0} }}} }{\frac{{1}}{{c\omega ^{2}}}}\sum\limits_{i}
n_{i} \left( {{\frac{{\hbar \omega}} {{\theta _{i}}} }} \right)^{1
/ 2}\times
\]
\begin{equation}
\label{eq58}\times\left\{ {{\frac{{\sin ^{2}\varphi _{i}}} {{m_{
\bot} \tau _{ \bot}  \left( {\theta _{i}}  \right)}}} +
{\frac{{\cos ^{2}\varphi _{i}}} {{m_{ \|} \tau _{\| } (\theta _{i}
)}}}} \right\} .
\end{equation}

From the comparison of formulas (\ref{eq49}) and (\ref{eq57}), one may see that the light
absorption coefficient in the classical frequency range depends equally on
the components of the tensor of relaxation times and the components of the
mass tensor both under impurity and acoustic scattering. The only difference
is in numerical coefficients, which is stipulated by the different energy
dependence of the relaxation times under impurity and acoustic scattering.

A totally different situation appears in the quantum region, as
can be seen from the comparison of Eqs. (\ref{eq53}) and
(\ref{eq58}). The reason for these differences is that the
impurity scattering potential (18) would have a singularity at
$\omega \to 0$ without screening (i. e., formally at $r_{\rm D}
\to \infty )$. Therefore, the screening effect for a charged
impurity should be taken into account in the classical region.
This screening is not significant in the quantum frequency region.
For the acoustic scattering, in difference to Eq. (\ref{eq51}), we
get
\begin{equation}
\label{eq59}
\mu _{\alpha}  = {\frac{{4}}{{3\sqrt {\pi}} } }{\frac{{e\tau _{\alpha}
\left( {\theta _{i}}  \right)}}{{m_{\alpha}} } }.
\end{equation}

\section{Polarization Effects under Light Emission by Free Carriers}

If the electron gas is heated (e.g., by electric current), the
effect opposite to the Drude absorption occurs, i.e., free
carriers emit light. Polarization dependences can appear in the
case of the anisotropic dispersion law of free carriers. Such
polarization effects take place in the cases of different heatings
or under the same heating but with different fillings of the
valleys.

We can obtain the spontaneous emission by hot electrons, which is
of interest for us, using the expression for the field-induced
emission, which we derived previously. For this, the vector
potential ($\vec {A}^{(0)})$ of a wave should be first normalized
so that $N_{\rm ph} $ photons are in volume $V$, i.e. the
following condition should be used:
\begin{equation}
\frac{1}{V}N_{\rm ph} \hbar \omega = \frac{E^{2}}{4\pi } =
\frac{1}{8\pi} \Biggl( \frac{\omega} {c} \Biggr)^{2}A^{(0)^{2}}.
\end{equation}
From here,
\begin{equation}
\label{eq60} A^{(0)} = 2c\left( {{\frac{{2\pi \hbar}} {{V\omega}}
}N_{\rm ph}}  \right)^{1 / 2}.
\end{equation}

Then we should substitute expression (65) in the formula for
$P^{(i)}( - )$, setting $N_{\rm ph} = 1$.

And, finally, we should multiply the obtained expression by the density of
finite field states in a unit frequency interval and a solid angle $d\Omega
$:
\begin{equation}
\label{eq61}
d\rho \left( {\omega}  \right) = {\frac{{V}}{{\left( {2\pi c}
\right)^{3}}}}\omega ^{2}d\Omega .
\end{equation}

As a result of procedures described, we obtain the following expression from
$P^{(i)}( - )$ for the emission of electrons in all the valleys into a solid
angle $d\Omega $ in the case of impurity scattering:
\[
 W^{( - )} = {\frac{{e_{0}^{6} n_{a} \sqrt {m_{ \|}}
d\Omega}} {{\left( {2\pi}  \right)^{3 / 2}\varepsilon _{0}^{2}
c^{3}\left( {m_{ \|}  - m_{ \bot}} \right)^{2}}}}\times
\]
\begin{equation}
\label{eq62}
 \times{\sum\limits_{i} {{\frac{{n_{i}}} {{\sqrt
{\theta _{i}}} } }}} e^{ - \hbar \omega / \theta _{i}}
{\int\limits_{0}^{\infty} {{\frac{{dxe^{ - x}{\left\{ {\Psi _{i}
(q_{\max}  ) + \Psi _{i} (q_{\min )} } \right\}}}}{{\sqrt {x(x +
\hbar \omega / \theta _{i} )}}} }}} .
\end{equation}

Expression (\ref{eq62}) gives the emission intensity from a unit volume with
${\sum\limits_{i} {n_{i}}}  $ electrons. To obtain the emission from an
arbitrary volume $V$, expression (\ref{eq62}) should be multiplied by $V$. Note that the
signs of the expressions $P^{(i)}( + )$ and $P^{(i)}( - )$ are different,
since $P^{(i)}( + )$ characterizes the energy incorporation into the
electron subsystem (i.e., absorption), while $P^{(i)}( - )$ describes the
energy extraction from it. In Eq. (\ref{eq62}), we use the absolute value of
emission intensity.

One may derive simple expressions in the limiting cases of the classical and
quantum frequency ranges from the general expression (\ref{eq62}), similarly to the
case of absorption.

In the case of the classical frequency range $(\hbar \omega < < \theta _{i}
)$, we have:
\[
W^{( - )} = {\frac{{e_{0}^{6} n_{a} \sqrt {m_{ \|}} } } {{\left(
{2\pi} \right)^{3 / 2}\varepsilon _{0}^{2} c^{3}\left( {m_{ \|} -
m_{ \bot}} \right)^{2}}}}\times
\]
\begin{equation}
\label{eq63}
 \times{\sum\limits_{i} {{\frac{{n_{i}}} {{\sqrt
{\theta _{i}}} } }} }\Psi _{i} (\infty )\ln \left( {C_{1}
x_{\min}}   \right)^{ - 1}d\Omega .
\end{equation}
We obtain, respectively, in the quantum frequency range $(\hbar
\omega > > \theta _{i} )$:
\[
 W^{( - )} = {\frac{{e_{0}^{6} n_{a} \sqrt {m_{ \|}}
} } {{\sqrt {2} \pi \varepsilon _{0}^{2}
c^{3}}}}{\frac{{1}}{{(m_{\|}  - m_{ \bot}
)^{2}}}}{\frac{{1}}{{\sqrt {\hbar \omega}} } }\times
\]
\begin{equation}
\label{eq64}
 \times{\sum\limits_{i} {n_{i}} }\Psi _{i} (\infty
)e^{ - \hbar \omega / \theta _{i}} d\Omega.
\end{equation}
Using the explicit expression (\ref{eq48}) for $\Psi _{i} (\infty
)$ and the formulas for the components of the relaxation tensor
under impurity scattering (\ref{eq50}), formula (\ref{eq63})
becomes
\[
W^{( - )} = {\frac{{3e_{0}^{2}}} {{16\pi ^{3 /
2}c^{3}}}}{\sum\limits_{i} {n_{i} \theta _{i} {\left\{
{{\frac{{\sin ^{2}\varphi _{i}}} {{m_{ \bot} \tau _{ \bot}  \left(
{\theta _{i}}  \right)}}} + {\frac{{\cos ^{2}\varphi _{i}}} {{m_{
\|}  \tau _{ \|}  (\theta _{i} )}}}} \right\}}d\Omega}}  .
\]
\begin{equation}
\label{eq65}
\end{equation}

The term $\ln (C_{1} x_{\min}  )^{ - 1}$ in Eq. (\ref{eq63}) is
related to the screening effect of the Coulomb potential of an
impurity. The screening effect is not significant in the quantum
frequency range. Therefore, this term does not appear in formula
(\ref{eq64}) which cannot be expressed by the components of the
relaxation tensor, as in the case of Eq. (\ref{eq65}).

Under the dominating role of acoustic scattering, the energy
emitted by all electrons in all the valleys per unit time into a
solid angle $d\Omega $ is equal to
\[
W^{( - )} = \frac{-2e_{0}^{2}} {3\pi ^{5 / 2}c^{3}}\sum\limits_{i}
n_{i} \theta _{i} e^{ - \hbar \omega / \theta _{i}}\times
\]
\[
 \times\Biggl\{\frac{\sin ^{2}\varphi _{i}} {m_{
\bot} \tau _{ \bot}( {\theta _{i}})} + \frac{\cos ^{2}\varphi
_{i}} {m_{ \|}  \tau _{ \|} (\theta _{i} )} \Biggr\} a_{i}^{3}
e^{a_{i}} \frac{d}{da_{i} }\left( {K}_{1} (a_{i} ) /{ a_{i}}
\right)d\Omega  .
\]
\begin{equation}
\label{eq66}
\end{equation}
We obtain from here for the classical frequency range ($\hbar
\omega < < \theta _{i} )$:
\[
W^{( - )} = {\frac{{4e_{0}^{2}}} {{3\pi ^{5 /
2}c^{3}}}}{\sum\limits_{i} {n_{i} \theta _{i} {\left\{
{{\frac{{\sin ^{2}\varphi _{i}}} {{m_{ \bot} \tau _{ \bot}  \left(
{\theta _{i}}  \right)}}} + {\frac{{\cos ^{2}\varphi _{i}}} {{m_{
\|}  \tau _{ \|}  (\theta _{i} )}}}} \right\}}d\Omega}}  .
\]
\begin{equation}
\label{eq67}
\end{equation}
A simple expression for the emission intensity can also be
obtained from Eq. (\ref{eq66}) in the quantum frequency range case
($\hbar \omega > > \theta _{i} )$:
\[
 W^{( - )} = {\frac{{e_{0}^{2}}} {{6\pi
^{2}c^{3}}}}\sum\limits_{i} {\frac{{n_{i}}} {{\sqrt {\theta _{i}}}
} }(\hbar \omega )^{3 / 2}e^{ - \hbar \omega / \theta _{i}}\times
\]
\begin{equation}
\label{eq68}
 \times \Biggl\{ \frac{\sin ^{2}\varphi _{i}} {m_{
\bot}  \tau _{ \bot} ( \theta _{i})} + \frac{\cos ^{2}\varphi
_{i}} {m_{ \|}  \tau _{ \|}  (\theta _{i} )} \Biggr\}d\Omega .
\end{equation}

We can see from Eq. (\ref{eq67}) that the emission intensity does
not depend on the emitted light frequency in the classical
frequency range, and drops exponentially in the quantum frequency
range. We recall that $\tau _{ \bot} $ and $\tau _{ \|}  $ in Eqs.
(\ref{eq66})---(\ref{eq68}) are the components of the acoustic
relaxation tensor, which are set by formula (\ref{eq54}).

We can see from the comparison of formulas (\ref{eq65}) and (\ref{eq67}) that the dependence
on the parameters is the same. The numerical coefficients are only
different, which is stipulated by the different dependences of the
relaxation tensor components on the electron energy under impurity and
acoustic scattering.

\section{Conclusion and Remarks}

In this paper, the general expressions are obtained for the
absorption coefficient, as well as for the emission intensity in the
presence of hot electrons. These expressions are derived with taking
into account the multivalley character of the electron spectrum as
well as the anisotropy of the dispersion law and the scattering
mechanisms. The obtained expressions depend both on the
concentration of electrons $n_{i} $ and their temperatures $\theta
_{i}$ in individual valleys.

In the case of thermodynamic equilibrium, all $\theta _{i} $
values are the same (and are equal to the lattice temperature).
Moreover, the populations $n_{i} $ in all valleys are also
identical when the unidirectional pressure is absent. Under
unidirectional pressure applied to a specimen, $n_{i} $ are the
known functions of the applied mechanical stress (e.g., see [5]).

Under an electric field applied, all $\theta _{i} $ and $n_{i} $
values (or a part of them) can be different. The procedure for
their calculation is well known (e. g., see [4]).

The especially simple case of the polarization dependence of
emission appears when all electrons migrate to a single valley. It
is remained the only dependence on one angle --- between the
polarization ort and a rotation axis of the ellipsoid of the
surface of equal energy of the populated valley. Experimentally,
the polarization dependences in \textit{n}-Ge have been
investigated in [6].

\rezume{%
╬╤╬┴╦╚┬╬╤╥▓ ╧╬├╦╚═└══▀ ▓ ┬╚╧╨╬╠▓═▐┬└══▀\\ ╤┬▓╥╦└ ┬▓╦▄═╚╠╚
┼╦┼╩╥╨╬═└╠╚ \\┬~~ ┴└├└╥╬─╬╦╚══╚╒\\ ═└╧▓┬╧╨╬┬▓─═╚╩└╒}{╧.╠. ╥юьўєъ}
{╬ЄЁшьрэю чруры№э│ тшЁрчш фы  ъюхЇ│Ў│║эЄр яюуышэрээ  ёт│Єыр
т│ы№эшьш эюё│ ьш │ │эЄхэёштэюёЄ│ ёяюэЄрээюую тшяЁюь│э■трээ  ёт│Єыр
урЁ ўшьш хыхъЄЁюэрьш т срурЄюфюышээшї эря│тяЁют│фэшърї. ╬ЄЁшьрэ│
тшЁрчш чрыхцрЄ№ т│ф ъюэЎхэЄЁрЎ│┐ хыхъЄЁюэ│т т юъЁхьшї фюышэрї │ ┐ї
ЄхьяхЁрЄєЁ. ┬Ёрїютрэю рэ│чюЄЁюя│■ чръюэє фшёяхЁё│┐ │ ьхїрэ│чь│т
Ёючё│ ээ  хыхъЄЁюэ│т. ╨ючуы эєЄю фюь│°ъютшщ │ ръєёЄшўэшщ ьхїрэ│чь
Ёючё│ ээ . ┬ёЄрэютыхэю яюы ЁшчрЎ│щэє чрыхцэ│ёЄ№ ёяюэЄрээюую
тшяЁюь│э■трээ  урЁ ўшї хыхъЄЁюэ│т. ╙ тшярфъє юфэюэряЁ ьыхэюую
Єшёъє рсю тхышъшї │эЄхэёштэюёЄхщ юяЁюь│эхээ  яюы ЁшчрЎ│щэє
чрыхцэ│ёЄ№ тш ты ║ │  ъюхЇ│Ў│║эЄ яюуышэрээ  ёт│Єыр т│ы№эшьш
хыхъЄЁюэрьш.}
\end{multicols}
\end{document}